 \documentclass[manuscript]{acmart}
\usepackage{multirow}
\usepackage{dirtytalk}

\AtBeginDocument{%
  \providecommand\BibTeX{{%
    \normalfont B\kern-0.5em{\scshape i\kern-0.25em b}\kern-0.8em\TeX}}}

\setcopyright{acmlicensed}
\copyrightyear{2024}
\acmYear{2024}
\acmDOI{XXXXXXX.XXXXXXX}

\acmConference[CUI '24]{CUI 2024 - 6th Conference on Conversational User Interfaces}{July 8-10, 2024}{Luxembourg City, Luxembourg}

\acmISBN{978-1-4503-XXXX-X/18/06}

\begin{document}

\title{Comparing Perceptions of Static and Adaptive Proactive Speech Agents}

\author{Justin Edwards}
\affiliation{%
  \institution{University of Oulu}
  \city{Oulu}
  \country{Finland}}
\email{justin.edwards@oulu.fi}

\author{Philip R. Doyle}
\affiliation{%
  \institution{IBM Research}
  \city{Dublin}
  \country{Ireland}}

\author{Holly P. Branigan}
\affiliation{%
  \institution{University of Edinburgh}
  \city{Edinburgh}
  \country{United Kingdom}
}

\author{Benjamin R. Cowan}
\affiliation{%
  \institution{University College Dublin}
  \city{Dublin}
  \country{Ireland}}
\email{benjamin.cowan@ucd.ie}

\renewcommand{\shortauthors}{Edwards, et al.}

\begin{abstract}
  A growing literature on speech interruptions describes how people interrupt one another with speech, but these behaviours have not yet been implemented in the design of artificial agents which interrupt. Perceptions of a prototype proactive speech agent which adapts its speech to both urgency and to the difficulty of the ongoing task it interrupts are compared against perceptions of a static proactive agent which does not. The study hypothesises that adaptive proactive speech modelled on human speech interruptions will lead to partner models which consider the proactive agent as a stronger conversational partner than a static agent, and that interruptions initiated by an adaptive agent will be judged as better timed and more appropriately asked. These hypotheses are all rejected however, as quantitative analysis reveals that participants view the adaptive agent as a poorer dialogue partner than the static agent and as less appropriate in the style it interrupts. Qualitative analysis sheds light on the source of this surprising finding, as participants see the adaptive agent as less socially appropriate and as less consistent in its interactions than the static agent. 
\end{abstract}



\maketitle

\section{Introduction}
As speech agents have become increasingly popular, users have highlighted multitasking during eyes-busy, hands-busy activities as a central motivation to trying these agents out \cite{luger_like_2016}. That said, users' initial excitement for speech agents is frequently diminished to the point of disappointment and even abandonment, owing to speech agent interactions falling short of their expectations in terms of their abilities as dialogue partners \cite{luger_like_2016, cowan2017infreq}. These expectations and internal models of speech agents as dialogue partners, termed \textit{partner models} \cite{branigan_role_2011,cowan_they_2017} play a key role in how speech agent users understand their interactions. People's interactions with speech agents may therefore benefit from greater alignment between agent behaviour and the expectations users have for them as nearly human-like dialogue partners \cite{cowan2017infreq, cassell2007body}. Indeed, research participants have unfavourably compared speech agents \cite{luger_like_2016} to human personal assistants, who they envision would be able to effortlessly help them multitask. In order for speech agents to meet these expectations, they will need to be able to interact proactively with users rather than waiting for the busy user to turn their attention to a speech interaction. 

 Insofar as a benefit of proactive speech agents would come from an ability to interact with users who are already engaged in another task, proactive speech interactions must sometimes begin while users are attending to something. That is to say, speech from proactive agents will sometimes interrupt attention to an ongoing task. Recent work has investigated the characteristics of human spoken interruptions of this type, and found that people take many of the same considerations that are mentioned in proactive agent design guidelines into account when interrupting another person - seeking to limit the distraction caused by their interruptions by limiting the duration of their speech, attempting to select good moments for their interruptions, and sometimes preceding their interruptions with access rituals to make them more socially appropriate \cite{edwards2021ElicitingSI,edwards_using_2023}. But these human interrupting behaviours have not yet been implemented in the design of speech agents. By combining the well-established design principles for proactive agent interactions and the recent descriptions of proactive human speech interactions, the present study aims to investigate the effect that designing a speech agent to be proactive will have on people's partner models of that agent as compared to their partner models of proactive speech agents which, like existing speech agents, do not adapt speech behaviours to a user's context. As this work represents the first to manipulate at adaptivity as an independent variable affecting perceptions of a proactive agent, adaptivity is manipulated broadly according to a variety of adaptive behaviours (explained in detail in section 3.3), seeking to establish evidence of an overall effect of adaptivity rather than isolating differential effects of particular adaptive behaviours.

\section{Related Work}
\subsection{Designing proactive agents}
 Some early work on agent based interaction sought to describe design principles for mixed-initiative agent-based interactions, sensitive to the principles which had guided the design of direct-manipulation user interfaces before them. Horvitz laid out 12 principles for mixed-initiative interfaces with this aim, including among others: considering uncertainty about a user’s goals, considering the status of a user’s attention in the timing of services, inferring ideal action in light of costs, benefits, and uncertainties, minimising the cost of poor guesses about action and timing, and employing socially appropriate behaviours for agent-user interaction \cite{horvitz_principles_1999}. Following from these early design principles, recent work on speech based proactive agents looked to test principles for the design of and implementation of proactive agents, with regards to details such as modality, timing, message content \cite{yorke-smith_design_2012, semmens_is_2019, cha_hello_2020}. The present study considers proactive agents which use speech to interrupt a user who is already engaged in a task. As such, it is necessary to further consider the design details of those specific types of proactive interactions. One study on the design of a learning assistant with these characteristics proposed nine principles for proactive agent behaviour, specifying that it should be valuable, pertinent, competent, unobtrusive, transparent, controllable, deferent, anticipatory, and safe \cite{yorke-smith_design_2012}. Echoing the general proactive agent design principles laid out by Horvitz \cite{horvitz_principles_1999}, this set of principles again focuses on adapting interactions based on contextual information, including contexts of the agent’s task, the user’s environment, and the social context of a non-human agent initiating interaction with a person.

Recent studies have demonstrated the extent to which people consider the context of the task they interrupt, including the difficulty and urgency of the task (in those works operationally defined as the interrupter's perceived cost of disrupting a dialogue partner) when forming their interruption \cite{edwards_using_2023, edwards2021ElicitingSI}. This work found that interruptions of urgent tasks are delivered more quickly, owing to adaptations to speech rate and word choice. It likewise found that people try to interrupt during relatively less demanding moments of an ongoing task and that they may forego politeness norms when the ongoing task is more difficult \cite{edwards_using_2023}. Other research in this area has focused on tone in interruptions, finding that assertive-voiced in-car notifications are less pleasant but more likely to elicit a response than a non-assertive voice \cite{wong_voices_2019}. The present study seeks to build on proactive agent design research by comparing user impressions of a proactive speech agent which adapts various characteristics of its speech according to the urgency and difficulty of an ongoing task against impressions of a proactive agent which ignores context and interacts in a static way. 

\subsection{Partner modelling of machine dialogue partners}
Speech agent interactions are a unique form of human-computer interaction as they require users to engage in dialogue with a machine dialogue partner, making the conversational abilities of that partner central to the interaction \cite{branigan_role_2011}. Prior research on spoken interactions, both those with people and with machines, have established the concept of partner models, the models by which people understand the capabilities of their dialogue partners \cite{cowan_voice_2015,branigan_role_2011}. Doyle and colleagues formally define partner models for machine dialogue partners as follows: \newline\emph{"The term partner model refers to an interlocutor’s cognitive representation of beliefs about their dialogue partner’s communicative ability. These perceptions are multidimensional and include judgements about cognitive, empathetic and/or functional capabilities of a dialogue partner. Initially informed by previous experience, assumptions and stereotypes, partner models are dynamically updated based on a dialogue partner’s behaviour and/or events during dialogue"} \cite{doyle_what_2021}

User studies of speech agent interactions have helped to establish that these partner models play a pivotal role in speech agent users’ overall experience of these interactions, with users finding interactions particularly unsatisfying when their experience does not match their partner model \cite{luger_like_2016, clark_what_2019}. In a qualitative study of users of popular speech agents like Siri and Google Assistant, users remarked on the promise of human-likeness, insinuated by marketing, human-like voice synthesis, and designed personalities which mimic a human personality, which created what the researchers called the “gulf of expectations” \cite{luger_like_2016}, following the more general “gulfs of execution and evaluation” described by Norman \cite{norman1983some}. Reflecting this research, it is critical for user experience that speech agents which prime human-like partner models to meet this expectation and deliver human-like capabilities.

Until recently, little research has explored the characteristics of partner models in speech agent interactions. Recent work has begun to investigate this question however, investigating the dimensions of partner models which are salient to people when engaged in dialogues with machines and with people \cite{doyle_dimensions_2022,doyle_mapping_2019, doyle_what_2021}. This research was further developed into a validated self-report questionnaire, the Partner Modelling Questionnaire (PMQ), across three factors (perceptions of partner competence and dependability, assessment of human-likeness, and perceptions of the communicative flexibility of the system) which can be used to measure people’s partner models for machine dialogue partners, indicating how strong of a dialogue partner a person views a given machine to be \cite{doyle_dimensions_2022}. By designing a proactive speech agent which adapts to a user’s context, this study aims to apply principles of proactive agent design to our current understanding of partner models in spoken interactions with machines. Specifically, this study aims to demonstrate that a proactive speech agent which adapts to a user’s context is perceived as more competent, human-like, and flexible than existing speech agents which are not adaptive to context. 

\subsection{Aims and hypotheses}
Prior research has highlighted the ways in which interruptions differ according to the urgency of the interruption and the complexity of task they interrupt. Holistically, these studies found that people adapt their interruptions in terms of timing, word choice, prosody, and the use of particular social markers (i.e. access rituals), taking urgency and task difficulty cues into account \cite{edwards_using_2023, edwards2021ElicitingSI}. This study aims to apply these findings to proactive non-human speech agents. Building upon those findings as well as research on the design of proactive agents \cite{horvitz_principles_1999} and partner modelling \cite{doyle_dimensions_2022}, the present study hypothesises the following:
\begin{itemize}
\item People will rate speech interruptions from an adaptive agent as coming at better moments as compared to interruptions from a static (non-adaptive) agent (H1)
\item People will rate speech interruptions from an adaptive agent as more appropriately asked as compared interruptions from a static (non-adaptive) agent. (H2),
\item People will view an adaptive agent as a stronger dialogue partner than their partner models for a static (non-adaptive) agent (H3). 
\end{itemize}

All hypotheses, research questions, and post-hoc analyses were pre-registered before data collection began.\footnote{osf.io/g8zk6/?view\_only=ec53ef395bd64ff3a13dae10e94775bb}

As in recent studies of interruptions of complex tasks \cite{edwards_using_2023, edwards2021ElicitingSI}, the study  uses Tetris as an ongoing, complex task which a proactive agent must interrupt with speech. Insofar as those studies provided a specific description of the ways people adapt their speech to contextual information in Tetris, the same task is used here so that the interruption adaptations observed in the human participants from those studies can be directly applied to the design of the proactive agent in this study. 

Rather than having participants act as Tetris players in this experiment, participants instead watched videos of interactions between the an agent and an unseen Tetris player. This video study technique is commonly used in human-robot interaction research \cite{ghafurian_2020_design,lohse_2008_evaluating} owing to its benefits in being rapidly deployable to many participants including online participants, greater standardised control over the interaction, and facilitation of the use of early-stage prototypes which may lack features necessary for live interactions \cite{woods_2006_video}. 


\section{Experimental Method}
\subsection{Participants}
80 crowdworkers (40 men, 40 women; M age = 38.4 years, SD = 11.9 years) were recruited on Prolific Academic. All participants were native speakers of English living in Ireland or the United Kingdom. 92.5\% (N=74) of participants reported having used speech assistants before, with 66.3\% (N=53) of participants reporting that they use a speech assistant once a week or more frequently. Participants were all familiar with Tetris, though most reported that they do not play frequently (81.3\% of participants answering 3 or lower on a 7-point Likert-type question asking “If you have played Tetris before, how often do you play Tetris?”). The study took approximately 20 minutes and participants were compensated £6 through Prolific Academic for their participation. The study received ethical approval through the university's ethics procedures for low-risk projects (Ethics code: [anonymized for review]).

\subsection{Materials}
Twenty-four videos of the game Tetris were created. In each video, a game of Tetris is played by an unseen player. Tetris videos were designed to match those described in prior work \cite{edwards_using_2023, edwards2021ElicitingSI}, which differentiated easy and hard games of Tetris to study how urgency affected people's interruptions in each context. Tetris videos from the present experiment were therefore either examples of easy games or hard games. Under the Tetris video, the word Urgent or Non-urgent appeared, indicating whether the video represented an urgent game of Tetris or a non-urgent game of Tetris, operationally defined as the interrupter's perceived cost of interrupting that game. In particular, participants in those studies were told that a Tetris player would rate their appropriateness and timing of their interruption, that on urgent trials, these ratings would contribute much more robustly to the interrupter's total score, and that their total score determined their chance at a cash bonus prize \cite{edwards_using_2023, edwards2021ElicitingSI}. In this way, urgent trials both in prior studies and the present research could be considered something more like "safety-critical" contexts, rather than something like "time-sensitive" contexts. 

A question was written at the bottom of each video, indicating the question that the proactive speech agent would be prompted to ask the Tetris player. After a fixed interval of 10 seconds, a large red dot indicator appeared in the video indicating that a proactive agent had been prompted to interrupt the player. The ten second delay was selected to give participants time to observe the Tetris game before an interruption might occur and reflects the maximum delay used in prior studies before prompting interruptions \cite{edwards_using_2023, edwards2021ElicitingSI}. After some delay (described below), a synthesised voice asked the question to the Tetris player. Videos end one second after the audio ends, with each video lasting approximately 20 seconds. 12 unique Tetris gameplay videos and prompts were sampled from \cite{edwards_using_2023}, resulting in 12 matched trials across two within-subjects conditions. 

Half of the videos selected (n = 6) were randomly sampled from the difficult Tetris games in \cite{edwards_using_2023}, in which videos started with a Tetris game piece at the top of the game board, at least half of the rows of the board which already contained Tetris pieces, and the falling speed of the game piece was set to 10 rows per second. The other half of the videos selected (n = 6) were randomly sampled from the easy Tetris games in  \cite{edwards_using_2023}, in which videos started with a Tetris game piece at the top of the game board, at least two rows and no more than half of the rows of the board already contained Tetris pieces, and the falling speed of the game piece was set to the game minimum of 1.25 rows per second.

For each difficulty grouping per condition, half of the videos (n = 3) are arbitrarily marked “urgent” and the other half (n = 3) are arbitrarily  marked “non-urgent”. Tetris gameplay, interruption prompts, and urgency are all fixed throughout each block of videos and across participants - for example, Tetris gameplay video 2, which came from the easy condition from  \cite{edwards_using_2023}, was urgent and had the prompt “What was the last movie you watched?” in both blocks for all participants. This controlled against unsystematic interactions between these variables, ensuring that differences experienced by participants are only the conditional differences described below. In keeping with prior studies, the content of a prompt is unrelated to both the urgency of the trial and the content of the Tetris task \cite{edwards_using_2023, edwards2021ElicitingSI}. 

\subsection{Experimental conditions- Adaptive vs Static Proactive Agent}
The experiment followed a one-way within-subjects design. Agent condition was manipulated across two conditions with respect to the timing of their interruptions, their speech rate, and the content of their interrupting utterance, all of which were fixed for the static agent and adaptive to context (with contexts varying in terms of urgency and game difficulty) for the adaptive agent. Details of the differences between agents are explained below.

\subsubsection{Adaptive agent}
Previous worked defined interruptible windows of Tetris games by classifying the characteristics of moments that viewers of Tetris games judged to be suitable for interruptions, in order to broadly label a variety of other Tetris games \cite{edwards_using_2023}. The adaptive agent varied its interruption onset and always began interruptions during one of these interruptible windows - typically moments with little active input from the player. Because prior work on human interruption adaptation manipulated task urgency and task difficulty as independent variables \cite{edwards_using_2023, edwards2021ElicitingSI}, the adaptations of the agent are in relation to the way that interruptions of urgent tasks differ from interruptions of non-urgent tasks and the way interruptions of easy tasks differ from the interruption of hard tasks.

For urgent interruptions, the adaptive agent interrupted at an onset three seconds after the red dot appeared, or as close as possible to three seconds while interrupting within an interruptible window. For non-urgent interruptions, the adaptive agent interrupted at an onset five seconds after the red dot appeared, or as close as possible to five seconds while interrupting within an interruptible window. The differences in interruption onsets was selected to reflect the difference in mean onsets observed in \cite{edwards_using_2023}, in which urgent interruptions came after a mean onset of 3.87  seconds whereas non-urgent interruptions came at a mean onset of 4.72 seconds. This difference was slightly exaggerated in the conditions presented in this experiment with the intention of making differences more salient to an observer. The use of interruptible windows by only the adaptive agent across all trials reflected prior findings in which participants did not significantly vary their usage of these windows by urgency or by Tetris difficulty conditions \cite{edwards_using_2023}.

The adaptive agent spoke at a 1.00 speech rate for non-urgent interruptions and a 1.10 speech rate for urgent interruptions, reflecting the difference in mean interruption durations observed in \cite{edwards_using_2023}, in which urgent interruptions lasted for a mean of 1519ms whereas non-urgent interruptions lasted for a mean of 1596ms. For all six of the trials which featured easy Tetris games, the adaptive agent used access rituals such as “hey” and “excuse me” to lead into the interruption. For the six trials which featured hard Tetris games, access rituals were not used, reflecting the difference observed in the use of access rituals in \cite{edwards_using_2023}, in which participants were significantly more likely to use access rituals during easy Tetris games. 

Finally, the adaptive agent rephrased all of its questions, using concise language (e.g. “Got any pets?” for the prompt “Do you have any cats or dogs?”) or conversational language (e,g, “Are you right or left-handed?” for the prompt “Which hand do you write with?”) for all trials, with these styles balanced across difficulty and urgency contexts. Each rephrased question was a verbatim recreation of the way a participant phrased the corresponding interruption from \cite{edwards_using_2023}. Concise language and conversational language were selected to reflect the two major rephrasing strategies mentioned in qualitative data in prior works, which found neither of phrasing styles to be exclusively associated with a single urgency condition \cite{edwards_using_2023, edwards_multitasking_2019}.  All interruption audio was synthesised using Google WaveNet text to speech. Half of the participants heard voice en-GB-Wavenet-A, a feminine voice, and the other half of participants heard voice en-GB-Wavenet-B, a masculine voice, fully balanced by participant gender.

\subsubsection{Static agent}
For the static agent, interruptions always began 4 seconds after the red dot appeared, or as close as possible while ensuring the interruption did not begin within an interruptible window as identified in \cite{edwards_using_2023}. The static agent asked questions exactly as they appeared on screen, with no changes to wording, at the standard 1.00 WaveNet speech rate, and without the use of any access rituals.

Overall, the static agent condition is meant to be representative of the capabilities of current speech agents like Google Assistant or Amazon Alexa, which do not use access rituals, vary speech rates, or vary the timing of their speech based on contextual cues. The adaptive agent was designed to adapt its speech in a variety of ways representative of the ways people were observed to adapt their speech in experimental studies. While this experimental design does not allow for the analysis of any particular type of adaptation’s causal relationship with interaction outcomes, it nonetheless gives a holistic representation of the overall effect of adaptation, directly informed by the approaches that people use to adapt speech for interruptions.

\subsection{Measures}
\subsubsection{Partner Model Questionnaire}
Participants were asked to complete the 18-item Partner Model Questionnaire (PMQ). The PMQ is a validated self-report scale consisting of word pairs separated by a 7-point semantic differential scale. The scale comprises three subscales: \textit{partner competence and dependability}, \textit{human-likeness}, and \textit{communicative flexibility} onto which nine, six, and three items load respectively \cite{doyle_dimensions_2022}. Scores are calculated for each subscale by summing semantic differential ratings for each word pair that loads onto the respective scale, with higher numbers corresponding to responses closer to the word more positively associated with that subscale (e.g. closer to the word "consistent" in the pair "consistent/inconsistent" which loads onto the \textit{partner competence and dependability} subscale). Total PMQ scores are calculated by summing the three component subscale scores. 

Participants were asked to complete the PMQ with the instructions "Thinking about the speech assistant you just watched, how would you rate its communicative ability on a scale between each of the following poles?". As a control, before the experiment began, participants were also asked to complete the PMQ in regards to the speech interface they are most familiar with. PMQ semantic differential item orders were randomised between participants, and 9 items were presented in reverse order (e.g. lower-scoring poles appeared on the left of the screen rather than the right) per participant, balanced across subscales. Items presented in reverse order were then reverse-scored so that presentation order did not affect scoring. 

\subsubsection{Single Item Questionnaires}
After each trial, participants were asked to answer on 5-point Likert-type scales how much they agreed with each of two statements: “The question came at a good moment” and "The assistant asked the question in an appropriate way." These items mirrored the themes described in \cite{edwards_using_2023}, timing and delivery, which participants identified as important features of the structure of a spoken interruption. 

\subsubsection{Demographic Questionnaire}
Participants were asked a number of questions about themselves including their age, nationality, level of expertise with Tetris, how recently they played Tetris, their level of experience with speech agents, and which speech agents they use.

\subsection{Procedure}
Participants were directed to a webpage where they read an information sheet describing the study and the data rights of participants. They were then asked to indicate their consent to participating in the experiment and sharing their anonymised data. Participants were told that they would watch 12 short videos of a person playing Tetris, during which the Tetris player would be interrupted by a proactive speech agent asking them a question. Participants were told that after each video, they would be asked to answer 2 questions about the interruption that they just watched and, after all 12 videos, they would be asked to complete a 18 item questionnaire about the agent they just listened to. The informational screens explained that after completing this routine once with one agent, they would then be asked to do the same again with a different agent. Participants were told that each agent was engaged in an exercise in which it needed to ask the Tetris player a variety of questions, and its goal was to minimise disruption to the Tetris player while asking its set of questions as quickly as possible. Informational screens explained that for some trials, minimising disruption to the player is urgent as the Tetris player was rated on their play during the game shown, rated games were used to choose the winner of a cash prize, and that the Tetris player did not know which games were rated.  

\begin{figure*}[h]
\centering
\includegraphics[width=.9\textwidth]{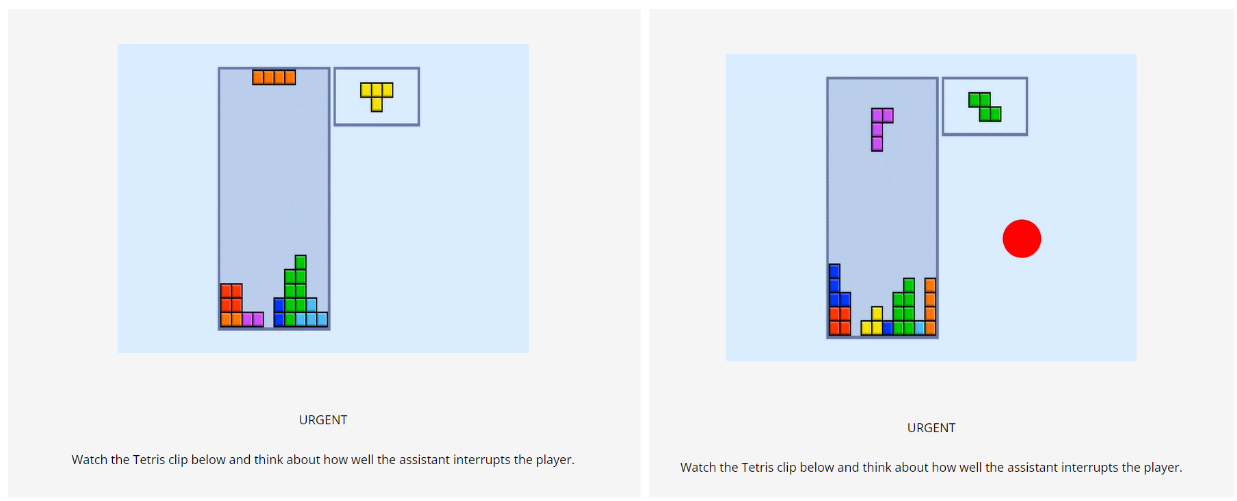}
\caption{Example screenshots from the experiment which participants saw as part of pre-test instructions. On the left, there is no red dot, so the agent has not yet been cued to interrupt. On the right, the red dot has appeared, signalling that the agent has been cued to interrupt.}
\label{fig:ch6-proc}
\end{figure*}

After this information was presented and the participants consented to take part in the study, they were asked to complete an initial PMQ questionnaire to get a baseline understanding of their views of speech agents in general. After the initial PMQ, participants saw example screenshots from a video. In the example screenshots, one image displayed a game of Tetris with a question prompt and the other screenshot displayed the same game of Tetris and question prompt with a large red dot overlaid above the Tetris game board (Figure \ref{fig:ch6-proc}). Participants were told that this visual indicator is not visible to the Tetris player, but it indicates to the observer (i.e. to the participant) that the agent has been prompted to ask a question to the Tetris player. Participants were informed that the agent could see the Tetris game and could decide when to begin its interruption any time after the interruption was prompted. After the participant viewed the example screenshots, they were shown a practice video in which a Tetris game is played by an unseen player, the visual indicator appears after some time, and a synthesised voice asks the Tetris player a question.

After the participant watched the practice video, they were asked to click a button to indicate that they were ready to continue and begin their first block of trials. Blocks of trials contained 12 videos of a single agent condition, with condition order counterbalanced across participants. Within a block of 12 trials, the order of videos was randomised for each participant. Following each video, participants rated how much they agreed with each of the following statements on a 5-point Likert scale: “The question came at a good moment” and “The question was asked in a disruptive way”. After each trial (video and Likert items), a plain white screen with a black central fixation cross appeared for a short interval before the next trial began. Each trial lasted between 10 and 20 seconds.

After completing the first block of trials, participants again completed an online version of the PMQ on a single webpage. After completing the PMQ, participants were asked to confirm that they were ready for the second block of 12 trials by clicking the continue button. After completing their second block of trials, participants completed another PMQ, reflecting the agent they just saw. 

After completing the final PMQ, participants were asked to complete a short demographic questionnaire. Participants were then thanked for their participation, given an opportunity to submit any other questions or comments, and debriefed on the aims of this study, including letting them know which block of trials was adaptive and which was the static agent. Finally, participants were given information for receiving payment. The full source code and materials for the experiment is provided\footnote{osf.io/g8zk6/?view\_only=ec53ef395bd64ff3a13dae10e94775bb}.

\section{Results}
\subsection{Analysis approach}
A total of 1920 interruption trials were viewed across the experiment by 80 participants, with participants responding to single-item questionnaires after each trial and to the Partner Model Questionnaire before the experiment and after each of the two agent conditions. Therefore, 1920 single-item questionnaire responses and 240 PMQ responses were recorded across all participants. No data needed to be removed for technical issues or by participant request. For PMQ responses, total and subscale scores within each condition were assessed for extreme values (+/- 3 standard deviations from the condition means) and none were detected. For each single item questionnaire, condition means were calculated for each participant and condition means were assessed across participants for extreme values (+/- 3 standard deviations from the between-participant condition mean) and none were detected. This resulted in all 1920 responses for each single-item questionnaire and all 240 full PMQ responses being included in the final analysis. 


Linear mixed-effects models were used to analyse the effect of agent condition on PMQ scores, single-item timing scores, and single-item appropriateness scores. Models were fit using the lme4 package version 1.1-26 \cite{bates_fitting_2015} in R version 4.1.1 \cite{r_core_team_r_2020}. Because PMQ responses were not measured for each video stimulus, the model of PMQ responses fits the fixed effect of agent condition (pretest, static, and adaptive) with intercepts per participant. The model of each single item questionnaire score fits fixed effects of agent condition (static and adaptive) with random by-participant and by-item slopes and intercepts (by-item effects include effects of stimulus, condition order, and trial order). Each model therefore represents the maximal model for that variable. Note that the urgency and Tetris difficulty of a given trial are not modelled individually as each stimulus is fixed in terms of Tetris clip (and thus Tetris difficulty and urgency condition). For PMQ models which have three levels of agent condition, the adaptive condition was used the reference level as H3 predicts PMQ differences between the adaptive and static conditions (but not differences between PMQ scores for either condition and the pretest scores). To improve reproducibility, full model syntax and random effect outputs are included for each model \cite{meteyard_best_2020}. Additional linear mixed-effects models were fit for each PMQ subscale as exploratory analysis to identify sources of differences between total PMQ scores. All analyses were preregistered before data collection began\footnote{osf.io/g8zk6/?view\_only=ec53ef395bd64ff3a13dae10e94775bb}.

\subsection{Quantitative response data}
\subsubsection{Single-item questionnaires}
\textit{Timing}:
For the first single-item questionnaire, “The assistant asked the question at a good moment”, there was no significant fixed effect of agent condition on participant ratings in a 5-point Likert-type scale [Unstandardised $\beta$ =-16, SE $\beta$ = .08, 95\% CI -0.26, 0.00], p = .053]. H1 is rejected. Full model syntax and output are included in Table \ref{tab:ch6-timing-model}. Means and standard deviations of single-item questionnaire responses by condition are presented in Table \ref{tab:ch6-single-desc}.

\begin{table}[]
\caption{Table of means and standard deviations for single-item questionnaire responses by condition}
\label{tab:ch6-single-desc}
\begin{tabular}{c|ccc}
\textbf{Measure} & \textbf{Condition} & \textbf{Mean} & \textbf{SD} \\ \hline
Timing & Static & 2.33 & 1.13 \\
 & Adaptive & 2.22 & 1.18 \\ \hline
Appropriateness & Static & 2.75 & 0.93 \\
 & Adaptive & 2.21 & 1.18
\end{tabular}
\end{table}

\begin{table}[]
\centering
\caption{Summary of fixed and random effects for timing single item questionnaire - Linear mixed effects model}  
Model: $ Timing \; rating = Agent \; Condition+ (1\vert subjectID)+(1+Condition\vert stimulus)+(1\vert trialOrder)$
\label{tab:ch6-timing-model}
\newline
\begin{tabular}{llllll}
\textbf{Fixed Effect} & \textbf{Std $\beta$} & \textbf{Unstd $\beta$} & \textbf{SE $\beta$} & \textbf{t} & \textbf{p} \\ \hline
Intercept & .07 & 2.35 & .14 & 16.83 & .001*** \\
Adaptive Agent & -.14 & -.16 & .08 & -2.10 & .053 \\
 &  &  &  &  &  \\
Random Effects &  &  &  &  &  \\ \hline
Group &  & SD & Corr &  &  \\ \hline
Participant (intercept) &  & .53 &  &  &  \\
Participant (slope) &  & .27 & -.08 &  &  \\
Stimulus (intercept) &  & .42 &  &  &  \\
Stimulus (slope) &  & .18 & -.47 &  &  \\
Trial order &  & .01 &  &  & 
\end{tabular}
\end{table}

\begin{figure*}[h]
\centering
\includegraphics[width=0.5\linewidth]{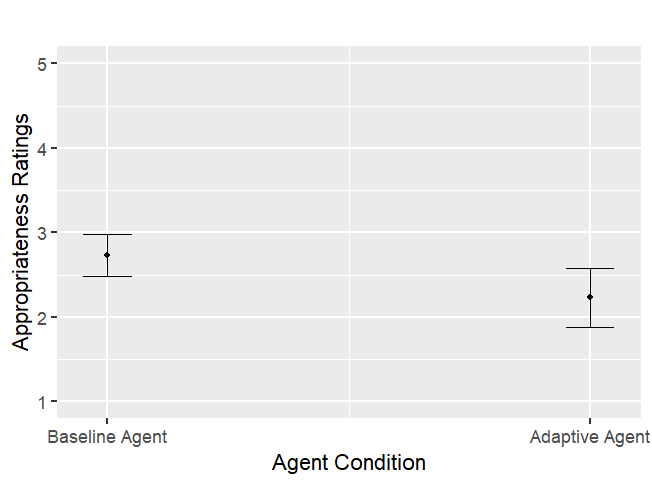}
\caption{Predicted values of appropriateness questionnaire ratings by condition}
\label{fig:ch6-approp}
\end{figure*}

\textit{Appropriateness}:
For the second single-item questionnaire, “The assistant asked the question in an appropriate way”, there was a significant fixed effect of agent condition on participant ratings in a 5-point Likert-type scale [Unstandardised $\beta$ =-0.58, SE $\beta$ = 0.17, 95\% CI -0.83, 0.-0.22], p = .003]. This indicates that participants rated questions asked by the static agent as being more appropriately asked than those asked by the adaptive agent. H2 is therefore rejected as the opposite result was found. This result is visualised in Figure \ref{fig:ch6-approp}. Full model syntax and output are included in Table \ref{tab:ch6-approp-model}.

\begin{table}[]
\centering
\caption{Summary of fixed and random effects for appropriateness single item questionnaire - Linear mixed effects model}  
Model: $ Appropriateness \; rating = Agent \; Condition+ (1+Condition\vert subjectID)+(1+Condition\vert stimulus)+(1\vert conditionOrder)$
\label{tab:ch6-approp-model}
\newline
\begin{tabular}{llllll}
\textbf{Fixed Effect} & \textbf{Std $\beta$} & \textbf{Unstd $\beta$} & \textbf{SE $\beta$} & \textbf{t} & \textbf{p} \\ \hline
Intercept & .26 & 2.75 & .14 & 20.33 & .002*** \\
Adaptive Agent & -.53 & -.58 & .17 & -3.40 & .003*** \\
 &  &  &  &  &  \\
Random Effects &  &  &  &  &  \\ \hline
Group &  & SD & Corr &  &  \\ \hline
Participant (intercept) &  & .72 &  &  &  \\
Participant (slope) &  & .66 & -.65 &  &  \\
Stimulus (intercept) &  & .08 &  &  &  \\
Stimulus (slope) &  & .51 & -.54 &  &  \\
Condition order &  & .76 &  &  & 
\end{tabular}
\end{table}

\subsection{Partner model questionnaire}
\begin{figure*}[h]
\centering
\includegraphics[width=0.5\linewidth]{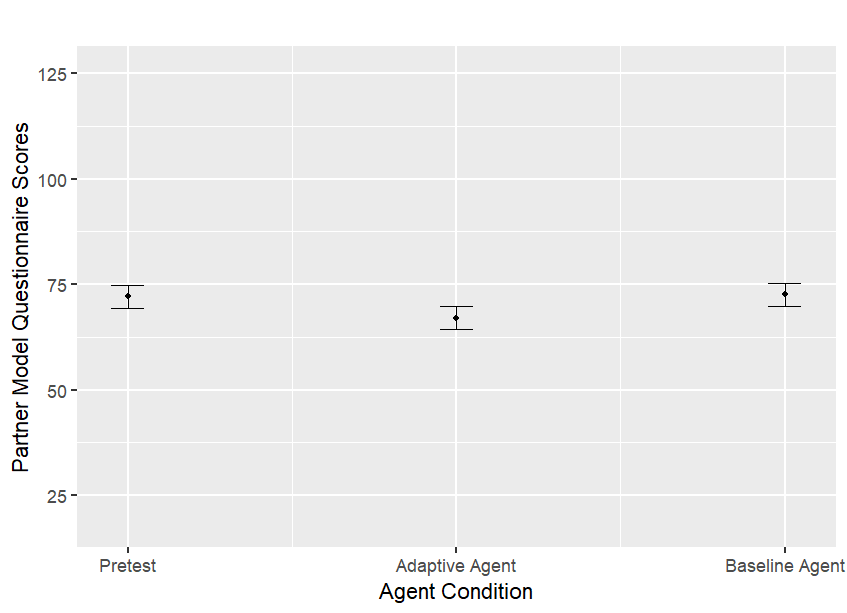}
\caption{Predicted values of Partner Model Questionnaire total scores by condition}
\label{fig:ch6-pmq}
\end{figure*}

There was a significant fixed effect of agent condition on Partner Model Questionnaire scores, with participants having significantly stronger partner models of speech agents before the experiment as compared with after interacting with the adaptive model [Unstandardised $\beta$ = 6.86, SE $\beta$ =  2.03, 95\% CI [2.86, 10.87], t = 3.38, p=.003] and stronger partner models of the static agent as compared to the adaptive agent [Unstandardised $\beta$ = 7.36, SE $\beta$ =  2.03, 95\% CI [3.36, 11.37], t = 3.63, p=.001]. H3 is therefore rejected as the opposite result was found, which is visualised in Figure \ref{fig:ch6-pmq}. PMQ and subscale means and standard deviations by condition are presented in Table \ref{fig:ch6-PMQ-desc}. Full model syntax and output are included in Table \ref{tab:ch6-pmq-model}. There was no difference between participants’ partner models of the static agent as compared with their pretest partner model of speech agents generally. This indicates that the manipulation was successful insofar as the static agent condition matched people’s prior impression of speech agents.

\begin{table}[]
\caption{Table of means and standard deviations for PMQ total score and subscale scores by condition}
\label{fig:ch6-PMQ-desc}
\begin{tabular}{c|ccc}
\textbf{Scale} & \textbf{Condition} & \textbf{Mean} & \textbf{SD} \\ \hline
\multirow{3}{*}{\begin{tabular}[c]{@{}c@{}}Total\\ PMQ\end{tabular}} & Pretest & 72.2 & 12.0 \\
 & Static & 72.6 & 11.1 \\
 & Adaptive & 67.0 & 13.9 \\ \hline
\multirow{3}{*}{\begin{tabular}[c]{@{}c@{}}Competence \&\\ Dependability\end{tabular}} & Pretest & 42.2 & 7.57 \\
 & Static & 42.8 & 6.66 \\
 & Adaptive & 37.3 & 8.96 \\ \hline
\multirow{3}{*}{\begin{tabular}[c]{@{}c@{}}Human-\\ Likeness\end{tabular}} & Pretest & 20.1 & 5.68 \\
 & Static & 19.8 & 6.54 \\
 & Adaptive & 19.9 & 6.95 \\ \hline
\multirow{3}{*}{\begin{tabular}[c]{@{}c@{}}Cognitive \\ Flexibility\end{tabular}} & Pretest & \multicolumn{1}{l}{9.90} & \multicolumn{1}{l}{2.74} \\
 & Static & \multicolumn{1}{l}{9.93} & \multicolumn{1}{l}{3.29} \\
 & Adaptive & \multicolumn{1}{l}{9.85} & \multicolumn{1}{l}{2.88}
\end{tabular}
\end{table}

\begin{table}[]
\centering
\caption{Summary of fixed and random effects for Partner Model Questionnaire total scores - Linear mixed effects model}  
Model: $ PMQ = Agent \; Condition+(1\vert subjectID)$
\label{tab:ch6-pmq-model}
\newline
\begin{tabular}{llllll}
\textbf{Fixed Effect} & \textbf{Std $\beta$} & \textbf{Unstd $\beta$} & \textbf{SE $\beta$} & \textbf{t} & \textbf{p} \\ \hline
Intercept & -.30 & 88.40 & 1.37 & 51.86 & <.001*** \\
Pretest & .44 & 6.86 & 2.03 & 3.38 & .003** \\
Static & .47 & 7.36 & 2.03 & 3.63 & .001** \\
 &  &  &  &  &  \\
Random Effects &  &  &  &  &  \\ \hline
Group &  & SD &  &  &  \\ \hline
Participant (intercept) &  & 8.22 &  &  & 
\end{tabular}
\end{table}

\begin{figure*}[h]
\centering
\includegraphics[width=0.5\linewidth]{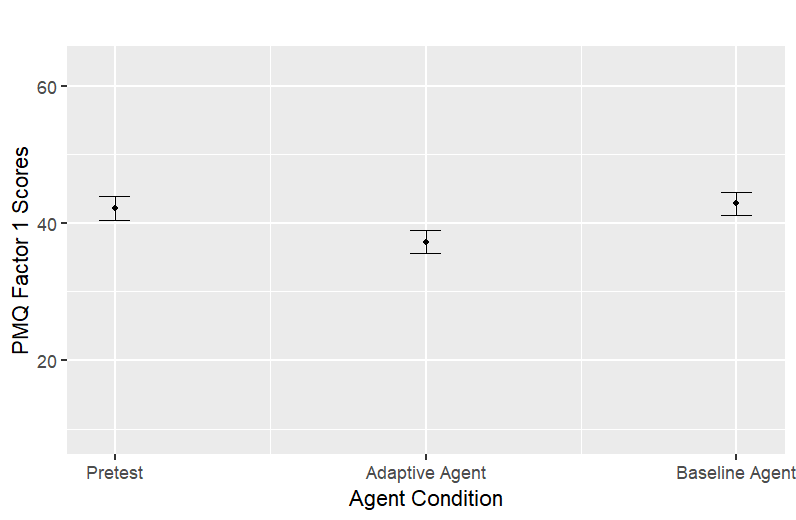}
\caption{Predicted values of Partner Model Questionnaire \textit{partner competence and dependability} subscale scores by condition}
\label{fig:ch6-pmq-f1}
\end{figure*}

To better understand the source of Partner Model Questionnaire differences between the agent conditions, further models were fit to compare participants’s scores across each of the three subscales of the PMQ. There was a significant fixed effect of agent condition on \emph{partner competence and dependability} subscale scores, with participants identifying speech agents as rating higher on this factor before the experiment as compared with after interacting with the adaptive model [Unstandardised $\beta$ =5.38, SE $\beta$ = 1.31, 95\% CI [2.79, 7.96]] and rating the static agent as stronger on this factor as compared to the adaptive agent [Unstandardised $\beta$ = 7.14, SE $\beta$ =  1.31, 95\% CI [4.55, 9.73]]. This result is visualised in Figure \ref{fig:ch6-pmq-f1}. There was no difference between participants’ \emph{partner competence and dependability} subscale ratings of the static agent as compared with their pretest partner model of speech agents. Full model syntax and output are included in Table \ref{tab:ch6-pmqf1-model}.

\begin{table}[]
\centering
\caption{Summary of fixed and random effects for Partner Model Questionnaire \textit{partner competence and dependability} subscale - Linear mixed effects model}  
Model: $ PMQ \; F1 = Agent \; Condition+(1\vert subjectID)$
\label{tab:ch6-pmqf1-model}
\newline
\begin{tabular}{llllll}
\textbf{Fixed Effect} & \textbf{Std $\beta$} & \textbf{Unstd $\beta$} & \textbf{SE $\beta$} & \textbf{t} & \textbf{p} \\ \hline
Intercept & -.41 & 51.64 & 1.09 & 47.38 & \textless{}.001*** \\
Pretest & .53 & 5.38 & 1.31 & 4.10 & \textless{}.001*** \\
Static & .70 & 5.44 & 1.31 & 5.44 & \textless{}.001*** \\
 &  &  &  &  &  \\
Random Effects &  &  &  &  &  \\ \hline
Group &  & SD &  &  &  \\ \hline
Participant (intercept) &  & 5.11 &  &  & 
\end{tabular}
\end{table}

There were no significant fixed effects of agent conditions on either \emph{human likeness} or \emph{conversational flexibility} subscales, indicating that overall PMQ differences between conditions are largely explained by differences in perceived competence and dependability.

\section{Discussion}
This study aimed to apply insights about human speech interruptions to the design of a proactive speech agent, investigating the effects of adapting speech to contexts of urgency and of ongoing task difficulty in the ways humans try to do when they interrupt. Prior work on speech agents has identified a gulf between user expectations and interaction realities \cite{luger_like_2016} owing to speech agents giving cues to users that they are more capable dialogue partners than they are revealed to be through interactions \cite{doyle_mapping_2019}. The present study therefore hypothesised that participants would rate speech interruptions from an adaptive agent as coming at better moments (H1) and as more appropriately asked (H2) as compared interruptions from a static agent. It further hypothesised that participants' partner models of an adaptive agent would be rated as stronger on the Partner Model Questionnaire \cite{doyle_dimensions_2022} than their partner model for the static agent (H3). 

These questions were investigated through an online experiment using prerecorded interactions between agent prototypes and a Tetris player. There was no significant difference between ratings of how well the agent timed its interruptions by condition, so H1 was rejected. Interruptions from the static agent were rated as statistically significantly more appropriately asked than those from the adaptive agent, so H2 was rejected. Likewise, participants' partner models of the proactive agent were statistically significantly weaker than their partner models of the static agent or their pretest control partner model of speech agents in general, as measured by the PMQ, so H3 was rejected. 

\subsection{Consistency as a salient feature for adaptive agents}
Contrary to expectation, the adaptive speech agent was rated lower on the PMQ by participants than was the static agent or people’s pretest perception of speech agents. Post-hoc analysis revealed that differences in PMQ scores resulted from differences in perceptions of partner competence and dependability. Reflecting on the items which load onto that PMQ factor, it becomes more clear why an adaptive agent would lead to a weaker partner model across this dimension. Items such as “Dependable/Unreliable”, “Consistent/Inconsistent”, and “Reliable/Uncertain” \cite{doyle_dimensions_2022} illustrate the importance of consistent, predictable behaviour in the formation of partner models. It may be the case that participants in this study did not have sufficient exposure to the adaptive agent to learn what contextual adaptations they could expect from the agent, leading to a poor understanding of those adaptations which cause them to seem arbitrary or inconsistent. 

As commercially available speech agents are not adaptive, participants’ mental models for the agents in this study would not likely lead to expectations of adaptivity.  Some research on adaptive interfaces has pointed toward the benefit of explicitly describing the sorts of adaptive features that an interfaces has and the errors that it may cause particularly for the purpose of setting appropriate expectations \cite{beggiato_evolution_2013}. Insofar as the novelty of adaptation diverges from prior experiences with speech agents, extended exposure to an adaptive agent or explicit coaching for mixed-initiative interactions may be beneficial in order to overcome perceptions of inconsistency. Returning to Horvitz’s principles of mixed-initiative interface design, agent behaviour should be socially appropriate \cite{horvitz_principles_1999}. While more research is needed to determine what people consider socially appropriate interactions between speech agents and people, it may be unsurprising when interactions with novel properties like contextual adaptivity are seen as socially inappropriate when they diverge from people’s prior experiences of similar interactions. Extended exposure and explicit descriptions of adaptive features should therefore be explored as ways of introducing potentially beneficial conversational features like adaptivity without introducing perceptions of inconsistency or inappropriateness.

Novel interactional elements may wane over time in the extent to which they are perceived as inconsistent. It is not clear how partner models develop over time with repeated exposure to a new partner, and the longitudinal work required to make that determination has been identified as a challenge for over a decade \cite{branigan_role_2011, cowan_voice_2015, doyle_dimensions_2022}. Even if prolonged exposure to adaptive proactive agent would improve people’s partner models of those agents, with better understanding of adaptive agents behaving consistently relative to particular contextual cues (rather than seeing them as inconsistent from utterance to utterance), this benefit is of little value when an early disappointment leads to abandonment of a system \cite{cowan2017infreq, luger_like_2016}. With sensitivity toward the negative impact of novelty on partner models and the effect of poor partner models on technological abandonment, further design of adaptive proactive speech agents may need to be more incremental than this study. Introducing adaptive features piecemeal across product lines or across time spent with an agent may be less jarring to a user than interacting with a speech agent with many novel design features introduced simultaneously. In order to make progress toward adaptive speech agents, the importance of consistent interactions must be considered.

\subsection{Appropriateness of adaptive proactive design}
The lack of improvement on PMQ scores for the adaptive agent as compared to static agent and pretest conditions was not only surprising because of a decrease in partner competence and dependability, but also because of a lack of increase in human-likeness. Similarly, while it was hypothesised that the adaptive agent would be rated as asking questions more appropriately, the opposite was found. Each of these findings can be better understood through the lens of appropriateness in human-machine dialogue. In addressing the gulf of expectations in speech agent interactions \cite{luger_like_2016}, some recent work has focused on the idea of appropriateness in these interactions \cite{aylett_right_2019, aylett_siri_2019, moore_appropriate_2017, le_maguer_synthesizing_2021}. This trend toward appropriateness has argued that increased human-likeness should not be a goal of itself in the design of speech agents. Instead, speech agents should be designed, in terms of voice \cite{aylett_siri_2019, le_maguer_synthesizing_2021} and in physical appearance in the case of embodied agents \cite{aylett_right_2019, moore_appropriate_2017}, to suit the role of the agent. Indeed qualitative work comparing human-human dialogue and human-machine dialogue has indicated that people see these two interactions as different in roles and in characteristics \cite{porcheron2018voice, reeves_conversation_2019}, with people expressing a dislike of speech agents which try to act human-like \cite{clark_what_2019, doyle_mapping_2019}. In this context, it may be clearer why participants rated the adaptive agent as no more human-like and as less appropriate in asking questions than the static agent. While adaptive interruptions may more appropriately utilise the context of an ongoing task, appropriateness also entails awareness of the social context.

\subsection{Individual differences and personalisation}
While neither the static nor the adaptive proactive agent were seen as significantly more human-like than participants’ prior conceptions of speech agents, this may be a result of differences between participants in opposing directions rather than a lack of difference in perceptions across participants. Research on personalisation of speech agents has found a high degree of variation between people in how they would like speech agents to be designed. Some strongly prefer agents which fulfil social functions, whilst others prefer agents to only perform non-social tool-like roles \cite{volkel_personalised_2020}. Likewise, prior research comparing the roles of conversations with machines to those with humans revealed tension between some people’s desires to have speech agents learn more about them to personalise interaction and others who saw speech agents building this sort of common ground with users as undesirable \cite{clark_what_2019}. Tailoring speech agent design to individual users may prove especially tricky due to high variance between individuals. Recent research on how people understand the personalities of speech agents found that the popular Big Five personality types used in human personality research proved less effective for classifying machines than a more graduated model of ten personality types \cite{volkel_developing_2020}. While more research is needed to determine differences among people’s preferences for these different personality types, it is clear that the design space for machine personalities is wide and that different designs are differentiable by the people that interact with these agents. These large individual differences in the perceptions of speech agent design decisions support the notion that personalisation of agents is both necessary and difficult. 

\subsection{Limitations}
Individuals in this study varied not only in their preferences toward speech agents, but also in their Tetris expertise. While this study mostly involved participants with some Tetris experience (i.e. neither experts nor total novices), there is sure to be variation in skill across participants. This may impact participants’ perceptions of the adaptive agent due to the differences in how expert and non-expert Tetris players perceive Tetris games \cite{lindstedt_distinguishing_2019}. Participants who have weaker understanding of Tetris gameplay and strategy were likely less sensitive to the state of the Tetris game and to the mental demand particular game states might put on the player. It may be for this reason that the adaptive agent’s consistent use of interruptible windows of Tetris for initiating its interruptions went unnoticed across the study, with participants finding neither agent as significantly better at timing its interruptions. Likewise, if the adaptive agent was not perceived as better at timing its interruptions, this may further help to explain why the adaptive agent was seen as less competent and dependable than the static agent. While prior research has demonstrated that people are somewhat skilful in identifying good moments to interrupt their own discrete tasks \cite{janssen_natural_2012}, their abilities to do so for a complex task like Tetris may be much more dependent on their expertise in that task. Future research should investigate both the effect of expertise on identifying interruptible moments in complex tasks.


Another limitation of the current study is the holistic manipulation of adaptivity rather than isolating particular adaptive behaviours or contexts for adaptation. While isolating individual behavioural or contextual variables would have allowed for a more precise description of causes to changes in participant perceptions of an agent, this study was the first to look at adaptivity as an independent variable in the design of a proactive speech agent. As such, there was little theoretical basis for choosing one finding over another when considering results from prior work which demonstrated the highly varied cues and decisions people consider when producing interrupting speech \cite{edwards_multitasking_2019}. This study thus presents an initial investigation into the salience and broad impact of adaptivity in this context without establishing particular causal links between particular behaviours and outcomes. Further work is needed to better understand these nuances such as the specific impact of access ritual use on perceptions of human likeness and the salience of using interruptible moments to deliver proactive speech during complex tasks. This study should be seen as an introduction to the question of how people want proactive agents to speak rather than a prescriptive set of design guidelines.

Furthermore, the PMQ has previously only been used to measure partner models formed by users of systems rather than observers who do not directly interact with the system, as in this study. As such, it may be the case that particular communicative features may be more or less salient to a person whose task was interrupted by a proactive agent as compared to an observer to the interaction. Given the level of control and more representative participant sample enabled by having participants observe an interaction rather than interacting directly, as well as the lack of adequate alternative tools for measuring partner model beliefs, we acknowledge that a limitation was introduced in the measurement of this variable.  Still, insofar as hypotheses were not only rejected but directionally backward from observed results, we do not believe the magnitude of differences in the formation of partner models between an interrupted person and an observer is likely to have materially changed the findings of this work. Indeed, as partner modelling research is still in its infancy, more work is needed to understand the cost of trading the control afforded by casting participants as experimental observers for the applicability of measuring the partner models formed by people who interact with a system directly. 

Finally, this study focuses solely on proactive speech interactions of a particular kind - personal questions - which interrupt a particular task - Tetris. These tasks were selected in order to maintain a high level of control over the structure of tasks and to maximally match the design of previous work in order to directly apply the findings of those studies to the design of this study. Salovaara and Oulasvirta describe the role of prototype experiments like this one as a way of evaluating possible futures \cite{salovaara_evaluation_2017}. In their framework, they describe design decisions as aimed at either staging - making the present have some characteristic of a possible future - or controlling - preventing particular characteristics of the present which are not expected to be part of the imagined future from becoming salient \cite{salovaara_evaluation_2017}.  While Tetris differs in a variety of ways from the sorts of tasks that people report wanting to use speech to multitask, like cooking or driving \cite{luger_like_2016}, it nonetheless matches the eyes-busy, hands-busy and complex, continuous nature of each of those tasks. In this way, the Tetris task maintains the aim of experimentally staging certain features of a possible future that this study represents. That said, it is not clear that proactive agents asking questions which are irrelevant to the user’s ongoing task represent a likely future use case for proactive systems. Nonetheless, the narrowing of speech tasks and stabilising of the tasks both across participants within this study serve the goal of controlling unsystematic variance, likewise helping to position this work to inform understanding of a possible future \cite{salovaara_evaluation_2017}. In order to explore different potential future interaction scenarios, further research will need to make carefully considered choices with regards to how to stage that future and control for unwanted present circumstances.

\section{Conclusion}
This study aimed to apply previous research insights about human spoken interruptions to the design of a proactive speech agent in order to assess people’s perceptions of such an agent. Applying prior work on the design of proactive non-human agents, this study identified adaptivity as a key variable for proactive agent interactions. As prior research has identified varied contextual cues that people consider when interrupting with speech and the different modifications they make in light of those contexts, this study manipulated adaptivity by designing one agent condition around those findings with the other, static agent condition that was insensitive to context. These agent conditions were compared across three measures: the PMQ, and single-item measures of how well interruptions were timed and how appropriately they were delivered. Quantitative results revealed that participants had a stronger partner model for the static agent as compared to the adaptive agent, owing to lower ratings of the adaptive agent’s competence and dependability. Participants likewise found the adaptive agent’s interruptions to be less appropriate than the static agent’s and detected no differences in the quality of the timing of interruptions between agents. These findings echo previous literature questioning the appropriateness of using human dialogue as a model or metaphor for nonhuman speech. While this early step toward the design of an adaptive, human-inspired, proactive speech agent revealed minimal overarching benefit to this approach, it may nonetheless serve as a guidepost for future investigation into this domain, by revealing tradeoffs encountered when using human behaviour to guide the design of speech agents.

\begin{acks}
This research was conducted with the financial support of Science Foundation Ireland under Grant Agreement No. 13/RC/2106\_P2 at the ADAPT SFI Research Centre at University College Dublin. ADAPT, the SFI Research Centre for AI-Driven Digital Content Technology, is funded by Science Foundation Ireland through the SFI Research Centres Programme. 
\end{acks}

\bibliographystyle{ACM-Reference-Format}
\bibliography{thesis}

\
\end{document}